Original Paper

Dylan Hamitouche[1,2]; Tihare Zamorano[1,3], BSc; Youcef Barkat[1,4], BSc; Deven Parekh[1], MSc; Lena Palaniyappan[1,5,6,7], MD, PhD; David Benrimoh[1,5], MD, MSc, MSc

[1] Douglas Mental Health University Institute

[2] Department of Medicine, McGill University

[3] Integrated Neuroscience Program (IPN), McGill University

[4] Department of Biochemistry, University of Montreal

[5] Department of Psychiatry, McGill University

[6] Robarts Research Institute, Western University, London, Ontario, Canada

[7] Department of Psychology, McGill University, Montreal, Quebec, Canada

Contributions:

DH helped conceive the study, performed the analyses, and wrote and revised the manuscript. TH, DP and YB helped complete the analyses, and reviewed the manuscript. LP provided funding and equipment and reviewed the manuscript. DB helped conceive the study and provided supervision.

# Sleep and Activity Patterns as Transdiagnostic Behavioral Biomarkers in Psychiatry: Initial Insights from the DeeP-DD study

Abstract

**Background:** Despite widespread use of symptom rating scales in psychiatry, these tools are limited by reliance on self-report, infrequent administration, and lack of predictive power. This constrains clinicians' ability to monitor illness trajectories or anticipate adverse outcomes like relapse. Actigraphy, a passive wearable-based method for measuring sleep and physical activity, offers objective, high-resolution behavioral data that may better reflect symptom fluctuations. Prior research has shown associations between actigraphy features and mood or psychosis symptoms, but most studies have focused on narrow diagnostic groups or fixed time windows, limiting clinical translation.

**Objective:** To examine whether actigraphy-derived sleep and activity features correlate with psychiatric symptom severity in a transdiagnostic psychiatric sample, and to identify which features are most clinically relevant across multiple temporal resolutions.

**Methods:** We present a feasibility case series study analyzing preliminary data from eight outpatients (ages 18–52) enrolled in the Deep Phenotyping and Digitalization at Douglas (DeeP-DD) study, a prospective transdiagnostic study of digital phenotyping. Participants wore wrist-based actigraphy devices (GENEActiv) for up to five months. Symptom severity was measured using a variety of self- and clinician-rated scales. We performed intra-individual Spearman correlations and inter-individual repeated measures correlations across daily, weekly, monthly, and full-duration averages. Longitudinal slopes of actigraphy and symptom trends were also analyzed.

**Results**: Intra-individual analyses revealed that later rise times were significantly associated with higher weekly PHQ-9 scores in participant #7 ($\rho = 0.74$, $P=.0003$) and participant #4 ($\rho = 0.78$, $P=.022$), as well as higher weekly GAD-7 scores in participant #7 ($\rho = 0.59$, $P=.026$). While similar trends were observed at daily and monthly timescales, the weekly resolution yielded the most robust significance. Inter-individual analyses showed that weeks with later average rise time correlated with higher PHQ-9 ($r = 0.48$, $P=.0003$) and GAD-7 scores ($r = 0.38$, $P=.032$), with the PHQ-9 association remaining significant after Bonferroni correction (Bonferroni-corrected $P=.015$). Increased light physical activity was linked to lower PHQ-9 scores weekly ($r = -0.44$, $P=.001$) and monthly ($r = -0.53$, $P=.014$). Over the whole duration of the study, increased levels of sedentary activity were associated with lower GAD-7 scores ($\rho=0.74$; $P=8.43 \times 10^{-23}$).

**Conclusion**:
Our findings highlight actigraphy-derived sleep and activity features, particularly rise time and physical activity, as promising transdiagnostic markers of psychiatric symptom burden. Their consistent associations across temporal scales and diagnostic groups underscore their potential utility for scalable, real-world clinical monitoring. Future work should validate these findings in larger cohorts and explore advanced analytical methods to capture circadian rhythmicity and symptom dynamics more precisely.

## 1. Introduction

Despite decades of research into the neurobiology of various mental illnesses, there remains a lack of clinically reliable tools for identifying, monitoring, and staging most psychiatric conditions[1-4]. In major depression, best practices involve standardized rating scales as part of measurement-based care, which can help patients achieve remission more quickly.[5-8] These scales are clinically useful for tracking symptom severity and guiding treatment decisions. However, there are no widely accepted protocols for monitoring patients post-remission, and in severe mental illnesses like schizophrenia, patient self-report is less reliable while rating scales add to clinicians' workload[9-12]. Moreover, rating scales may fail to capture what matters most to patients[13]—for instance, the commonly used PHQ-9[14,15] does not distinguish between hypersomnia and insomnia. While rating scales are helpful for assessing illness status or guiding treatment adjustments, they are seldom used to predict events like rehospitalization. This may be due in part to their limited predictive utility, as well as their infrequent and inconsistent use in clinical settings. Therefore, this gap in our ability to anticipate and prevent adverse outcomes highlights the need for more reliable, actionable predictors in clinical care.

In response, and in parallel with the rise of smartphones and wearables, interest has grown in "digital biomarkers"[16,17]—sensor-derived data collected passively by patient devices. These digital biomarkers often use both smartphones and wearables, capturing data such as app usage, communication, location, light, movement, sleep, activity, and heart rate.[18,19]. These tools have been used to infer mood states or behavior by combining multiple data streams—e.g., call logs and GPS for sociability and mobility[19]. Integrating passive (sensor) and active (questionnaire) digital tools with clinical, historical, and biological data is one approach to "deep phenotyping"—a detailed characterization of patients. Among these tools, actigraphy, a wearable-based method for measuring rest–activity cycles, has emerged as particularly relevant. It provides objective, continuous data on sleep and activity, which are critical in conditions like bipolar disorder[20,21] and are often poorly captured by self-report[24]

Actigraphy has shown clinical relevance by detecting changes linked to symptom trajectories across psychiatric disorders. In schizophrenia, sleep disturbances captured by actigraphy may worsen symptoms via mood and attention-related pathways.[23-25] Studies show daily fluctuations in sleep and activity track symptoms, with physical activity linked to improved same-day mood but sometimes worse next-day symptoms.[26] In mood disorders, machine learning models using actigraphy can predict next-day depressive or manic episodes.[27] Depressed patients show reduced daytime activity, with age-specific sleep difficulties suggesting circadian disruption as a core mechanism.[28-30] Finally, in bipolar disorder, sleep and activity disturbances persist even during euthymia and are tied to episode onset, severity, and long-term outcomes.[31–33]

Despite their promise, existing studies of actigraphy-derived sleep and activity features have often examined narrow diagnostic groups or short, fixed observation windows, limiting generalizability.[34,35] This limitation is significant given the multitude of available actigraphy measures, making it essential to identify which features demonstrate reliability and clinical relevance across diagnostic categories to optimize data collection and enhance clinical interpretation. Furthermore, since actigraphy data can be examined over multiple temporal scales, from days to months, understanding which measures maintain their utility across these timescales is critical. For instance, measures that are only informative over extended periods restrict their applicability for early or real-time clinical inference, as a substantial accumulation of data would be necessary before meaningful conclusions can be drawn. Addressing these limitations will improve the clinical utility and efficiency of actigraphy-based monitoring in psychiatric research and clinical practice.

This gap in the literature motivates the exploration of actigraphy-derived sleep and physical activity features as candidate biomarkers that transcend diagnostic categories and temporal scales. As part of the Deep Phenotyping and Digitalization at Douglas (DeeP-DD) project, a prospective transdiagnostic study aiming to identify clinically useful, scalable, and interpretable digital markers of mental illness, we present a case series study aiming to demonstrate the feasibility of actigraphy-based monitoring across diagnostic categories and temporal scales. We report data completeness rates, highlight the relevance of actigraphy for tracking subjective mood within and between individuals, and explore early signals of clinical relevance across multiple timescales in a real-world, diagnostically heterogeneous sample. Identifying robust, objective markers of illness could support clinical psychiatry by reducing reliance on disorder-specific tools and providing utility even when diagnostic clarity is limited, such as in youth mental health populations.

## 2. Methods

Deep Phenotyping and Digitalization at Douglas (DeeP-DD) is an ongoing feasibility study at the Douglas Mental Health University Institute in Montreal, Canada. Its aim is to explore digital phenotyping in a realistic transdiagnostic population including individuals with early psychosis, schizophrenia/schizoaffective disorder, bipolar disorder, anxiety, unipolar depression, and other conditions (e.g., personality or substance use disorders). Participants are encouraged to engage at a level that is feasible for them, and are compensated for their time; this compensation scales with the number of datastreams patients engage in to reflect the extra time required. The study aims to identify optimal measures and sampling frequencies across data types and diagnoses, and to develop useful, interpretable clinical reports to support personalized care and recovery. In this article, we focus on preliminary actigraphy and questionnaire data from a pilot phase of the study as these were the data types available for the most patients.

2.1 Participant recruitment

Participants are outpatients recruited from the Clinical High-Risk for Psychosis Clinic, First-Episode Psychosis Clinic, and the Bipolar Disorder Clinic; one patient with a primary personality disorder diagnosis was recruited from the general Neuropsychiatry Clinic. Participants are referred to the study by their clinicians, or self-refer to the study after seeing study advertisements at their clinic. Informed consent was obtained from all participants and those who agreed to provide actigraphy data were provided with a GENEActiv wristband (Activinsights, Cambridge, United Kingdom) and given instructions on its use. Participants received monetary compensation for wearing the watch and completing questionnaires. The study was approved by the research ethics board of the West-Central Montreal health authority and conducted in line with the Declaration of Helsinki and the Tri-Council Policy Statement.

2.2 Data collection

Patient- and clinician-rated questionnaire data were collected using the REDCap platform[36]. For every participant, a medical chart review was performed at the Douglas Mental Institute assessing overall symptomatology, medication use, and other clinically relevant information. Actigraphy data was extracted from the GENEActiv wristbands using GENEActiv PC Software (ver. 3.3)[37], a validated approach to monitor sleep in adults.[31–33] Sleep and physical activity features extraction were processed from the raw movement data using GENEActiv default R markdown analysis tools[38]. Participants wore the actigraphy device for one month at a time (the length of a single charge), after which it was replaced with a newly charged device. The sleep features extracted and included in the analyses were: total sleep time, sleep efficiency (i.e. time spent asleep divided by time spent in bed), number of active periods per night, median length of those active periods, sleep onset time, rise time, day-to-day sleep onset time variability, and day-to-day rise time variability. The GENEActiv PC Software provided activity modes classified as "sedentary", "light physical activity", "moderate physical activity", and "vigorous physical activity".[37] We excluded vigorous activity from our analyses, as participants spent very little time in this mode (often less than 5 minutes per day), and these brief episodes were likely due to artifacts or misclassification rather than true vigorous exertion. The daily number of steps was also provided by the model. The REDCap questionnaires included in the analyses were the Patient Health Questionnaire 9 (PHQ-9)[14], which assesses depressive symptoms; Generalized Anxiety Disorder 7 (GAD-7)[39], which evaluates anxiety severity; Clinical Global Impression - Severity (CGI-S)[40], which measures overall illness severity; Scale for the Assessment of Positive Symptoms (SAPS)[41], which examines psychotic symptoms; and Scale for the Assessment of Negative Symptoms (SANS)[42], which assesses deficits in normal emotional and behavioral functioning. The SAPS, SANS, and CGI-S were clinician-rated, while the PHQ-9 and GAD-7 were self-reported. The PHQ-9 and GAD-7 were administered weekly, the SAPS and SANS monthly, and the CGI-S at each clinical visit. Actigraphy data were inaccessible to participants and clinicians during the monitoring month but were made available to both after each monthly monitoring period concluded. As this was a real-world clinical feasibility study, both the timing of clinician-administered questionnaires and return of results depended on clinical follow-up schedules and as such were subject to variation in timing.

2.3 Statistical analyses

To characterize both individual-level variability and group-level patterns, we performed intra- and inter-participant analyses across multiple time scales. Features were averaged by time unit (daily, weekly, monthly and overall duration of study) to enable consistent comparisons between behavioral data and psychiatric outcomes. Due to the limited sample size, missing data were not imputed to avoid introducing bias.

We conducted analyses at the intra-participant level to capture how actigraphy features relate to psychiatric symptoms within each individual over time. For each participant, we computed Spearman rank correlations between actigraphy features and questionnaire scores at daily, weekly, and monthly resolutions, but only if at least five data points were available to ensure sufficient data for a stable and interpretable correlation estimate. This non-parametric method was chosen due to its robustness to non-linear associations and ordinal or skewed data distributions. Correlations were computed independently for each participant and time scale. Bonferroni correction was applied across all tests to control for multiple comparisons.

Then, we performed analyses at the inter-participant level to identify consistent associations across the sample, accounting for repeated measures within individuals.[43] To do so, we used repeated measures correlation (Pingouin package, Python)[44] to assess associations between daily, weekly, and monthly actigraphy features and questionnaire scores across individuals, while accounting for non-independence of within-subject data. Additionally, we computed Spearman correlations across participants using data aggregated over the full study period to capture stable between-subject associations and overall group-level trends not dependent on repeated measurements. To ensure reliable estimation, only correlations with at least 10 data points for daily analyses and 4 for weekly and full-study summaries were included, as inter-individual analyses provided more data than intra-individual ones. These thresholds were chosen to balance data availability and reliability.

To assess long-term co-evolution between symptoms and actigraphy-derived features over time, we computed the slope of each actigraphy metric and questionnaire score over the whole duration of the study using linear regression. Spearman correlations were then performed across participants between these slopes for each actigraphy–symptom pair. False discovery rate (FDR) correction was applied to account for multiple comparisons. This approach emphasizes directional consistency rather than absolute values, aiming to detect digital behavioral markers that co-vary with symptom trajectories.

## 3. Results

### 3.1 Description

Data were collected from eight participants, yielding a combined total of 33 months of actigraphy recordings. Participants were aged 18 to 30, included both males and females, and represented three primary psychiatric diagnoses with various comorbidities. (Table 1) The mean number of days of actigraphy data collected per participant was 110 (SD=43), ranging from 58 to 158 days. A total of 57 days were missing due to sensor malfunction, distributed across three participants (#5, #7, and #14). Data missingness is reported in Table 2. Most participants met criteria for a first episode of psychosis (FEP), except for participant #5 who had a primary personality disorder and participant #14 who had a primary bipolar disorder. Primary diagnoses included schizophrenia spectrum disorders (e.g., schizophrenia, schizoaffective disorder), bipolar disorder, and personality disorder. Comorbid conditions commonly included depression, anxiety, ADHD, PTSD, and personality disorders. Participants differed in clinical severity and care utilization: Participants #4 and #8 had histories of hospitalization and multiple ER visits, while others (e.g., participants #1, #5, and #13) had no acute care history. Medication use ranged from none (patient #1) to complex regimens involving antipsychotics, mood stabilizers, antidepressants, and stimulants (e.g., participants #5 and #8) (*Table 1*). Participants #1 and #6 did not consent to complete self-reported questionnaires.

**Table 1.** Clinical and sociodemographic characteristics of participants.[a-u]

| Participant | Age | Sex | Employed | Clinic | Primary diagnosis | Diagnostic impression | Psychiatric Family History | Hospitalizations | ER Visits |
|---|---|---|---|---|---|---|---|---|---|
| #1 | 20's | F | No | CHR | CHR-P | CHR-P, Dissociative identity disorder, Complex trauma, Anxiety, potentially transferring to psychosis | MDD, Anxiety, BPD, SUD | 0 | 0 |
| #4 | 20's | M | No | FEP | SSD | FEP in the form of mania, most likely Schizoaffective disorder given negative symptoms, r/o concomitant mood disorder | Schizophrenia, Anxiety | 1 | 1 |
| #5 | 30's | M | Yes | Neuropsychiatry clinic | Mixed PD | Mixed personality disorder (clusters B and C), Gambling addiction | None | 0 | 0 |
| #6 | 20's | F | Yes | FEP | SSD | Schizophrenia,, Depression, Social anxiety | None | 0 | 1 |
| #7 | 20's | M | Yes | FEP | BD-1 with psychotic features | BD-1, ADHD, Mixed anxiety dis (GAD/SAD), SUD (cannabis), C-PTSD with dissociation, Panic attacks, Functional neurological disorder, r/o: personality dis (mixed with paranoia, narcissistic traits, | None | 0 | 1 |

| | | | | | | OCDP), r/o: tourettes, r/o: OCD | | | |
|---|---|---|---|---|---|---|---|---|---|
| #8 | 10's | M | No | FEP | SSD + ASD | Autism, Schizophrenia, MDD, likely gender dysphoria, ASD, Anxiety NOS, r/o ADHD | MDD, SAD, Schizophrenia, Bipolar Disorder | 1 | 3 |
| #13 | 20's | F | No | FEP | SSD | Psychosis NOS, likely schizoaffective-depressive type, Mixed anxiety disorder (GAD, panic attacks), Tourette's syndrome (not currently, major focus of treatment), OCD traits, ADHD, Auditory processing | Schizophrenia | 0 | 0 |
| #14 | 50's | F | Yes | Bipolar clinic | BD-2 | BD-II, Anxiety disorder | Bipolar disorder, Depression, Anxiety disorder, Gambling addiction | 0 | 2 |

[a]Age was rounded down to the nearest decade to preserve confidentiality.
[b]ADHD: Attention-Deficit/Hyperactivity Disorder
[c]ASD: Autism Spectrum Disorder
[d]BD-I: Bipolar Disorder type I
[e]BD-II: Bipolar Disorder type II
[f]BPD: Borderline Personality Disorder
[g]C-PTSD: Complex Post-Traumatic Stress Disorder
[h]CHR: Clinical High Risk
[i]CHR-P: Clinical High Risk for Psychosis with Positive symptoms
[j]FEP: First-Episode Psychosis
[k]GAD: Generalized Anxiety Disorder
[l]MDD: Major Depressive Disorder

[m]NOS: Not Otherwise Specified
[n]OCD: Obsessive-Compulsive Disorder
[o]OCDP: Obsessive-Compulsive Personality Disorder
[p]PD: Personality Disorder
[q]PTSD: Post-Traumatic Stress Disorder
[r]r/o: Rule out
[s]SAD: Social Anxiety Disorder
[t]SSD: Schizophrenia Spectrum Disorder
[u]SUD: Substance Use Disorder

**Table 2.** Summary of actigraphy adherence and questionnaire completion

|  | Days of Actigraphy | Missing Days of Actigraphy | % Completed Actigraphy | PHQ-9 Completed | GAD-7 Completed | CGI-S Completed | SAPS/SANS Completed |
| --- | --- | --- | --- | --- | --- | --- | --- |
| Mean ± SD | 110.1 ± 40.4 | 7.1 ± 11.5 | 95.4 ± 7.5 | 9.8 ± 11.0 | 7.5 ± 9.5 | 2.8 ± 1.2 | 2.1 ± 0.8 |
| Min | 58 | 0 | 78 | 0 | 0 | 0 | 1 |
| Max | 156 | 35 | 100 | 29 | 25 | 4 | 3 |

## 3.2 Intra-participant patterns

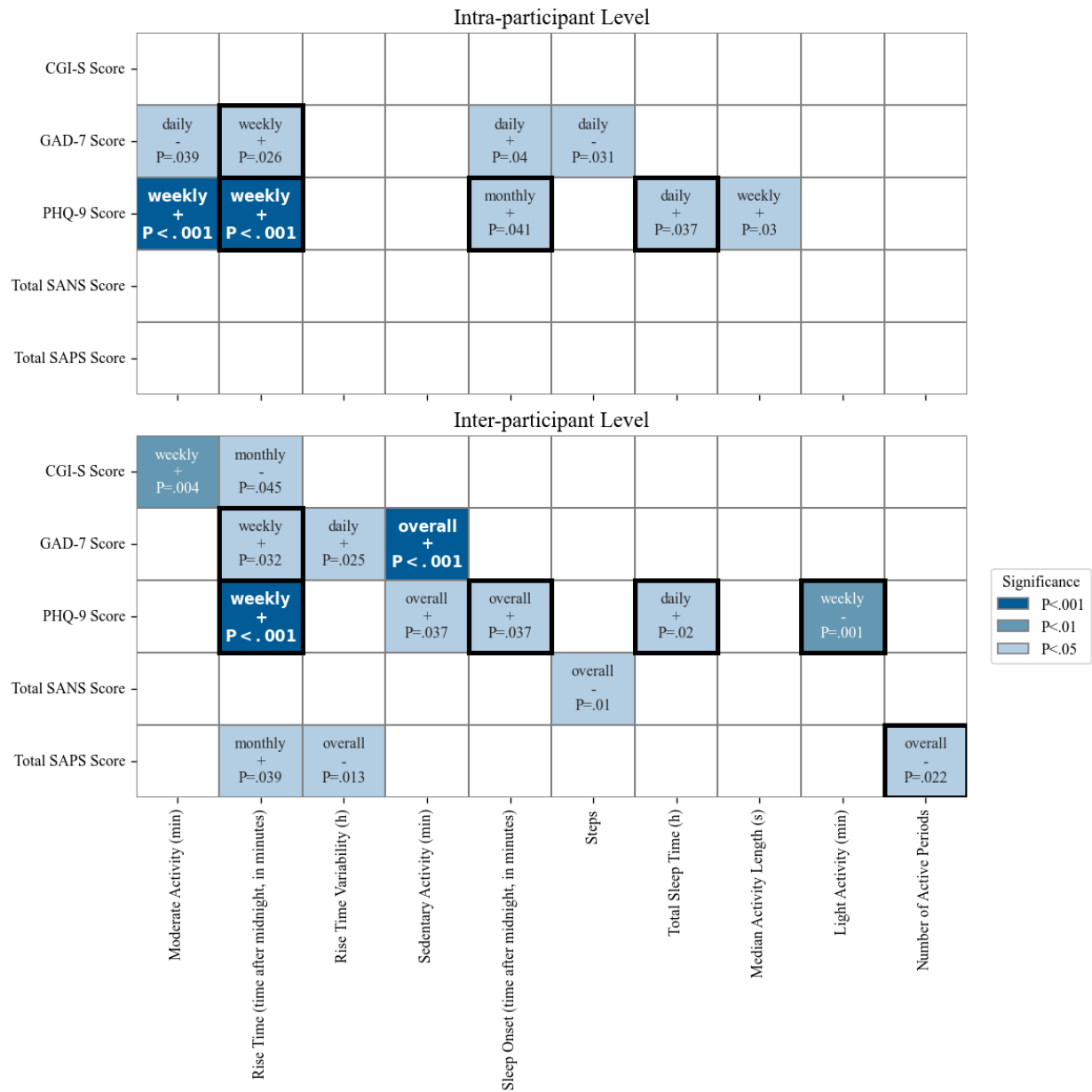

**Figure 1.** Heatmap of significant associations between actigraphy features and clinical questionnaires across multiple time scales at both intra- and inter-participant levels. "+" and "−" indicate the direction of the correlation. Cells with bold borders represent features with significant associations at multiple time scales (and/or across different participants for intra-individual analyses). Each cell displays only the most significant time scale.

### 3.2.1 Daily

We analyzed same-day associations between questionnaire scores and actigraphy features within participants using Spearman correlations. In participant #7, higher GAD-7 scores were associated with a later sleep onset the night before (ρ = 0.54, *P*=.040) and later same-day rise time (ρ = 0.53, *P*=.041). However, these daily associations did not remain significant after Bonferroni correction. *Figure 1* summarizes all significant associations established in this study.

*3.2.2 Weekly*

On a weekly time scale, we found that for participant #4 (patient with a schizoaffective disorder and a personality disorder), a later rise time was associated with higher PHQ-9 scores, although the significance did not survive Bonferroni-correction (ρ=0.8; uncorrected *P*=.016 Bonferroni-corrected *P*=.177). For participant #7, (patient with bipolar disorder, personality disorder, and a first-episode psychosis) a later rise time was associated with both higher PHQ-9 (ρ=0.74; *P*=.0003) and GAD-7 scores (ρ=0.59; *P*=.026), although only the association with PHQ-9 stayed significant after correction (*P*=.007). We found that for participant #5 (patient who has a primary personality disorder), longer time of moderate physical activity was associated with higher PHQ-9 scores (ρ=1.0; *P*=1.83x10$^{-23}$); see discussion below. This association correlates clinically, as it was noted in his clinical chart that he tended to go on long walks in the evening when feeling depressed. The higher level of significance for this participant is explained by the perfect correlation and limited number of overlapping data points for this participant (5 weeks), which can inflate correlation coefficients.

*3.3.3 Monthly*

Participant #7 showed that later average monthly rise time was associated with higher monthly average GAD-7 score (ρ=0.81; *P*=.05), while later average monthly sleep onset was associated with higher monthly average PHQ-9 scores (ρ=0.77; *P*=.041). None of these correlations stayed significant after Bonferroni correction.

In summary, several associations emerged between mood scores and sleep timing or physical activity. After correction, only weekly associations involving rise time and PHQ-9, and moderate activity and PHQ-9 (which was most likely idiosyncratic to participant #5), remained significant—highlighting rise time and activity patterns as potential behavioral markers of mood.

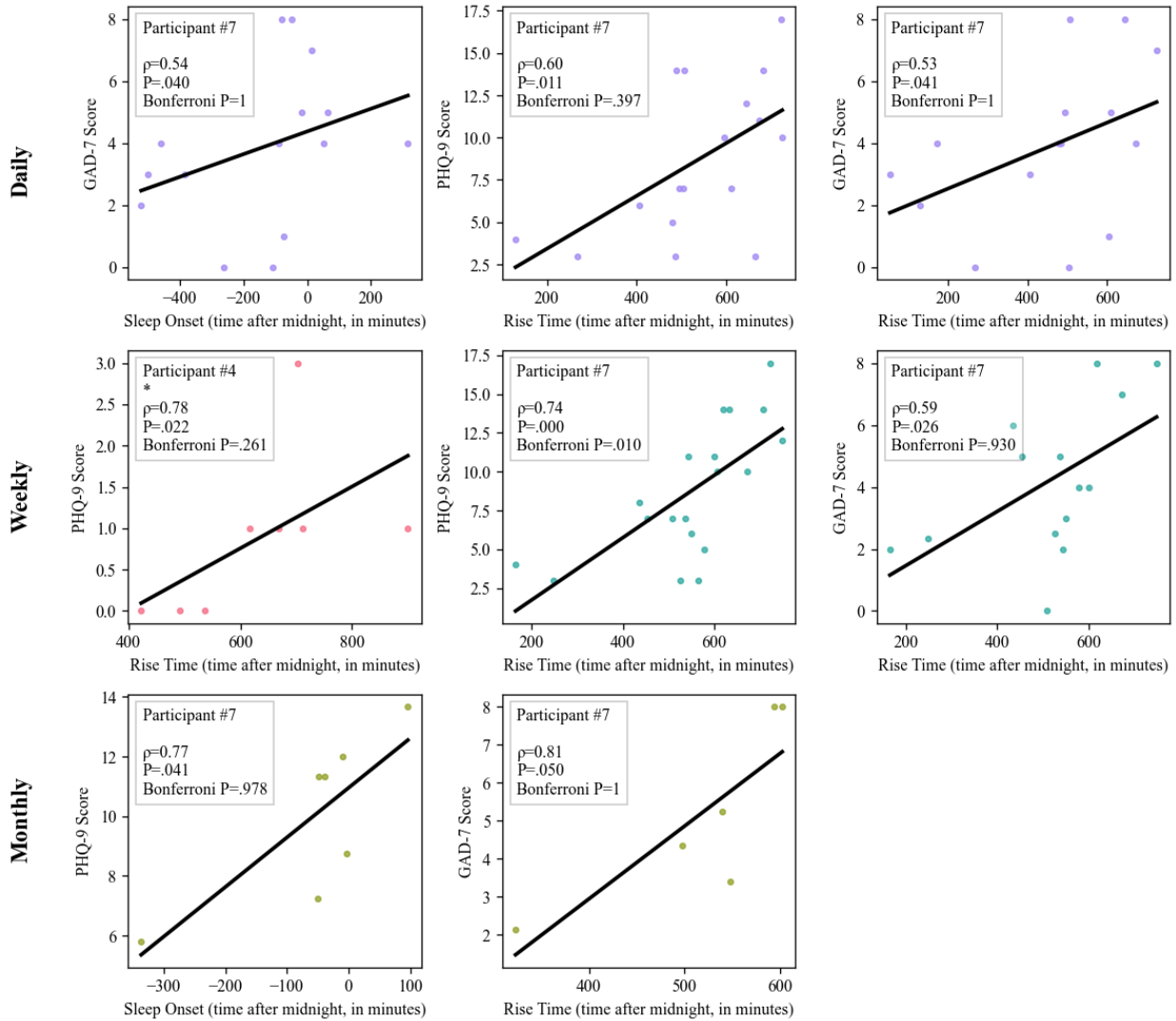

**Figure 2.** Significant Intra-Individual Associations Between Actigraphy Features and Questionnaire Scores Across Time Scales. * indicates fewer than 10 data points; Spearman correlation may be less reliable.[45]

3.4 Inter-participant trends

*3.4.1 Daily*

Same-day associations at the inter-individual level showed that longer total sleep duration the night before (r=0.30; *P*=.029). Increased rise-time variability, defined as the absolute difference in rise time between consecutive days, capturing day-to-day fluctuations in wake-up time, was correlated with higher GAD-7 scores (r=0.39, *P*=.019). However, none of the same-day associations survived correction for multiple comparisons.

*3.4.2 Weekly*

Interestingly, the intra-participant associations between rise time, sleep onset time, and questionnaire scores stayed consistent in our inter-participant repeated measures correlation analyses conducted at a weekly time scale, such that weeks with a later average rise time were associated with a higher GAD-7 (r=0.38; *P*=.032) and PHQ-9 scores (r=0.48; *P*=.0003). Furthermore, weeks of increased time spent doing light physical activity were associated with lower PHQ-9 (r=-0.44; *P*=.001). Only the association between PHQ-9 scores and rise time stayed significant after Bonferroni correction (*P*=.016).

*3.4.3 Monthly*

On a monthly time scale, increased light physical activity was associated with lower PHQ-9 scores (r=-0.53; *P*=.014). Later average monthly rise time was associated with both higher average CGI-S (r=-0.68; *P*=.045) and lower total SAPS (r=0.96; *P*=.039). The association with higher CGI-S appears contradictory to the rest of the study's findings linking later rise time with better outcomes, but this is likely driven by participant clustering—specifically, participant #4, who had consistently later rise times but low CGI-S scores. However, none of the monthly associations stayed significant after Bonferroni correction.

*3.4.4 Overall duration of study*

When comparing participants' average questionnaire scores and actigraphy features collected over the study period (ranging from one to several months, depending on each participant's enrollment date), Spearman correlation analyses showed that participants with a later average sleep onset had a higher PHQ-9 scores average ($\rho$=0.90; *P*=.037). Participants with more average daily time spent in sedentary activity had higher PHQ-9 scores ($\rho$=0.90; *P*=.037) and GAD-7 scores averages ($\rho$=1.00; *P*=1.40x10$^{-24}$); however, after Bonferroni correction, only the GAD-7 association remained significant (*P*=8.43×10$^{-23}$), though this unusually strong correlation should be interpreted with caution given the small sample size (see *Figure 3*).

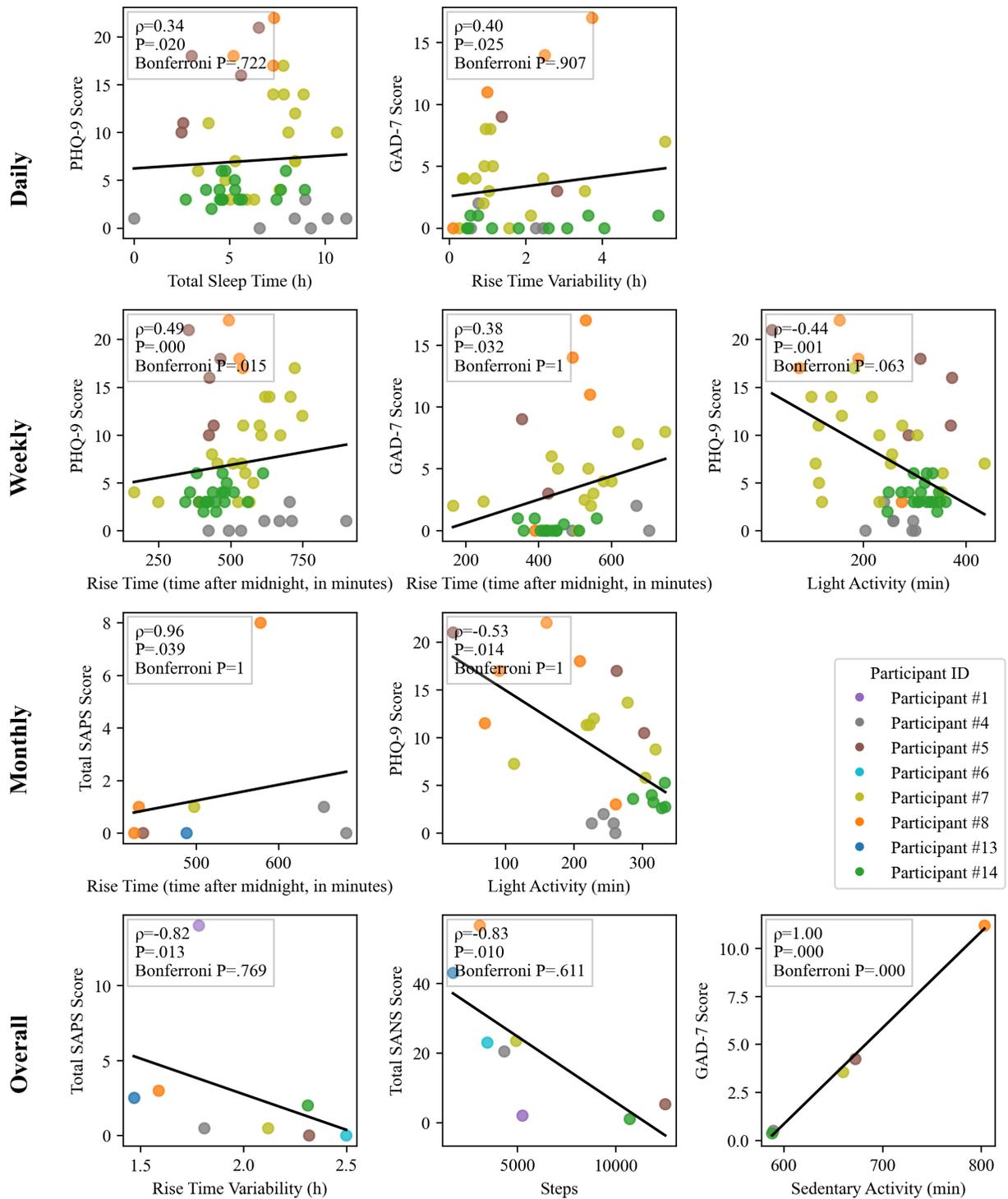

**Figure 3.** Significant Inter-Individual Associations Between Actigraphy Features and Questionnaire Scores Across Time Scales. Repeated measures correlations were conducted at the daily, weekly, and monthly time scales; Spearman correlations were used for overall

averages across the full study duration. * indicates fewer than 10 data points; Spearman correlation may be less reliable.[45]

To explore potential longitudinal trends between actigraphy features and symptom trajectories, we computed slopes of actigraphy metrics and symptoms over the duration of the study, then correlated slopes using Spearman correlation for each actigraphy–symptom pair across participants. Decreasing PHQ-9 scores over time was associated with increasing sleep efficiency (r = -0.94, $P$=.005), increasing daily time of light physical activity (r = -0.94, $P$=.005), and with decreasing sedentary behavior (r = 0.83, $P$=.04). Increasing GAD-7 scores over time correlated with delayed sleep onset over time (r = 0.94, $P$=.005), while increasing CGI-S scores associated with decreasing daily time of light physical activity (r = -0.82, $P$=.03). None of these associations remained significant after FDR correction. Full regression results and additional findings are provided in the *Multimedia Appendix*.

In summary, findings from inter-participant analyses aligned with intra-participant results, showing consistent trends where symptom improvement was linked to earlier rise times, longer sleep duration, and increased time spent doing light physical activity, while symptom worsening was associated with increased sedentary behavior and delayed sleep onset. These associations were stronger at the weekly level, with some remaining significant after correction for multiple comparisons.

## 4. Discussion

This case series, focused on the feasibility of the use of actigraphy in a realistic clinical setting, explored the relationship between sleep and activity patterns and psychiatric symptom severity across multiple diagnostic categories and time scales. Despite our small and heterogeneous sample, several meaningful intra- and inter-individual associations emerged. Notably, sleep timing (particularly rise time) showed consistent associations with mood scores at the individual level, but also at the group level and at different time scales. While these results align with prior evidence linking delayed sleep-wake cycles to mood disorders and psychosis, our study is novel in showing these associations within a realistic transdiagnostic population and highlighting their persistence across different temporal scales.[40,41] This holds clinical significance, as it underscores the potential of sleep timing as a modifiable biomarker and intervention target for improving mood symptoms across a range of psychiatric disorders.

At the individual level, higher depressive and anxious symptoms were consistently associated with delayed circadian rhythm, observed at both daily and weekly time scales. For participant #5, whom we know clinically to become more agitated when depressed, the presence of worse depressive symptoms during weeks of increased moderate physical activity likely reflects this pattern. This highlights the importance of accounting for individual clinical context when interpreting behavioral data. Inter-individual analyses echoed these findings, showing that weeks with more severe depressive and anxious symptoms were also characterized by later rise times. The persistence of these associations after correction in this small, heterogeneous sample strengthens the case for delayed sleep–wake phase (i.e. a regular sleep schedule that is considerably later than the conventional or desired time) as a transdiagnostic marker of psychiatric symptom burden. This characteristic has been previously linked with worse depressive symptoms in young adults[48] and worse outcomes in patients at clinical high risk of psychosis, as well as in those with early psychosis or schizophrenia[47,49], but we provide new evidence that these significant changes may occur across temporal scales and diagnostic groups[50,51]. We also found that during weeks and months of increased physical activity, participants reported less severe depressive and anxiety symptoms, consistent with findings from previous studies.[52]

Same-day analyses, although not surviving correction for multiple comparisons, showed that shorter total sleep duration the night before, earlier rise time and a rise time that has less variability from the previous day are each associated with lower same-day symptom severity. The unexpected association between total shorter sleep duration the preceding night and lower same-day symptom severity should be interpreted with caution, as the relationship was weak and largely driven by a small number of participants (notably participants #5 and #7). The sleep timing findings further support prior evidence that circadian phase is a key predictor of same-day mood episodes.[26] These trends point toward possible short-term responsiveness of mood and anxiety symptoms to daily behavioral patterns.

Additionally, when comparing participants' data across the entire study period (i.e. averaged across all months), those with later average sleep onset and more sedentary behavior tended to report worse symptoms, while those with higher average daily step counts reported fewer negative psychotic symptoms. Longitudinal analyses suggest that improvements in psychiatric symptoms, especially depressive symptoms, are accompanied by earlier sleep onset, increased sleep efficiency and increased levels of physical activity. This may reflect a delay correction or phase advance in circadian rhythms as mental health improves, supporting once again the idea that sleep timing and physical activity could serve as a sensitive marker of symptom trajectory across both short- and long-term time scales.

Despite the promising findings described above, several limitations must be acknowledged. The small sample, as well as missing data due to sensor malfunctions and variable adherence to actigraphy and questionnaire protocols, may have limited statistical power, introduced bias and reduced sensitivity to detect certain effects. Furthermore, it is important to note that inferring temporal causality from these observations remains challenging; incorporating Ecological Momentary Assessments (EMA) data could provide a more precise temporal resolution, thereby improving the ability to disentangle the directionality of the relationship between sleep timing and symptom trajectory.[53] Nonetheless, the consistency of results across time scales and methods highlights sleep timing and physical activity as a promising behavioral biomarker in real-world psychiatric populations.

## 5. Conclusion

The preliminary findings from the DeeP-DD study reveal complex associations between sleep features and clinical symptoms across various diagnostic groups, time scales, and data collection methods within a realistic clinical population. Earlier sleep and wake times, along with higher physical activity levels, were consistently associated with better clinical outcomes, including lower anxiety and depressive symptoms, both within and between individuals and across multiple time scales. The weekly time scale was particularly interesting, as multiple associations involving both sleep timing and physical activity were significant across participants. Although some associations did not remain significant after correction for multiple comparisons, the consistent patterns observed at both the inter- and intra-individual levels align with prior literature and offer novel insights into which actigraphy metrics, for different durations, may best predict clinical trajectories in a transdiagnostic population. The next steps in our project involve further exploring these relationships using larger datasets to assess their consistency and investigate the longitudinal dynamics between sleep and symptom changes. Additionally, we aim to incorporate more sophisticated approaches such as Fourier transformations at an intra-individual level to capture rhythmicity and periodic patterns in sleep and activity data, which may reveal subtle disruptions in circadian cycles linked to symptom fluctuations.[54] These methods could help identify individual-specific signatures, ultimately informing personalized interventions.

## Acknowledgements

The authors acknowledge financial support from the Mach-Gaensslen Foundation of Canada, the Dr. Clarke K. McLeod Memorial Scholarship, the Saputo Foundation, the Emerging Challenges Modelling Project (an initiative led by the Centre de Recherches Mathématiques in partnership with GERAD and UNIQUE, and funded by the Quebec Research Fund), the Strategia Program at the Centre de Recherches Mathématiques, the McGill Computational and Data Systems Initiative, the Douglas Research Center, the Fonds de recherche du Québec – Nature et technologies (FRQNT) (STRATÉGIA grant), and the Fonds de Recherche du Québec - Santé (FRQS) (Junior 1 Grant).

DB is supported by a NARSAD Young Investigator Award. LP's research is supported by Monique H. Bourgeois Chair in Developmental Disorders, the Graham Boeckh Foundation, the Mirella and Lino Saputo Foundation, and a Wellcome Trust Discretionary Grant (226168/Z/22/Z to Dr. Iris Sommer and LP). He receives a salary award from the Fonds de recherche du Québec-Santé (FRQS: [366934](#)).

## Conflicts of Interest

Dr. David Benrimoh is a founder and shareholder of Aifred Health, a digital mental health company. Aifred Health was not involved in this research. LP reports personal fees for serving as chief editor from the Canadian Medical Association Journals, speaker honorarium from Janssen Canada and Otsuka Canada, SPMM Course Limited, UK; book royalties from Oxford University Press; investigator-initiated educational grants from Otsuka Canada outside the submitted work, in the last 5 years.

## Data Availability

Data available on reasonable request from the authors and subject to review board approval.

## Abbreviations

ADHD: Attention-Deficit/Hyperactivity Disorder
ASD: Autism Spectrum Disorder
BD-I: Bipolar Disorder type I
BD-II: Bipolar Disorder type II
BPD: Borderline Personality Disorder
C-PTSD: Complex Post-Traumatic Stress Disorder
CHR: Clinical High Risk
CHR-P: Clinical High Risk for Psychosis with Positive symptoms
CGI-S: Clinical Global Impression - Severity
DeeP-DD: Deep Phenotyping and Digitalization at Douglas
EMA: Ecological Momentary Assessment
FDR: False Discovery Rate
FEP: First-Episode Psychosis
GAD: Generalized Anxiety Disorder
GAD-7: Generalized Anxiety Disorder 7
MDD: Major Depressive Disorder
NOS: Not Otherwise Specified

OCD: Obsessive-Compulsive Disorder

OCDP: Obsessive-Compulsive Personality Disorder

PD: Personality Disorder

PHQ-9: Patient Health Questionnaire 9

PTSD: Post-Traumatic Stress Disorder

r/o: Rule out

REDCap: Research Electronic Data Capture

SAD: Social Anxiety Disorder

SANS: Scale for the Assessment of Negative Symptoms

SAPS: Scale for the Assessment of Positive Symptoms

SSD: Schizophrenia Spectrum Disorder

SUD: Substance Use Disorder

## Multimedia Appendix

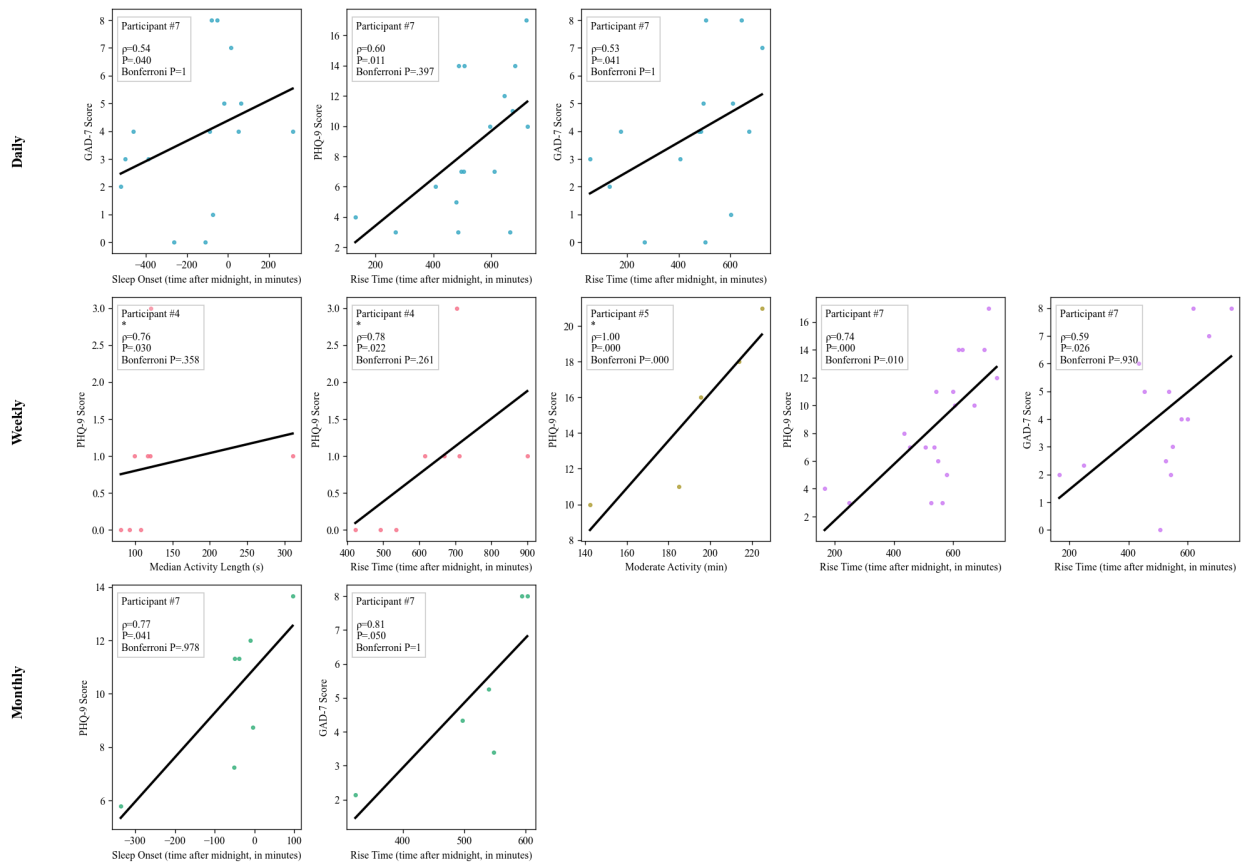

**Figure 4.** Complete Set of Significant Intra-Individual Associations Between Actigraphy Features and Questionnaire Scores Across Time Scales. * indicates fewer than 10 data points; Spearman correlation may be less reliable.

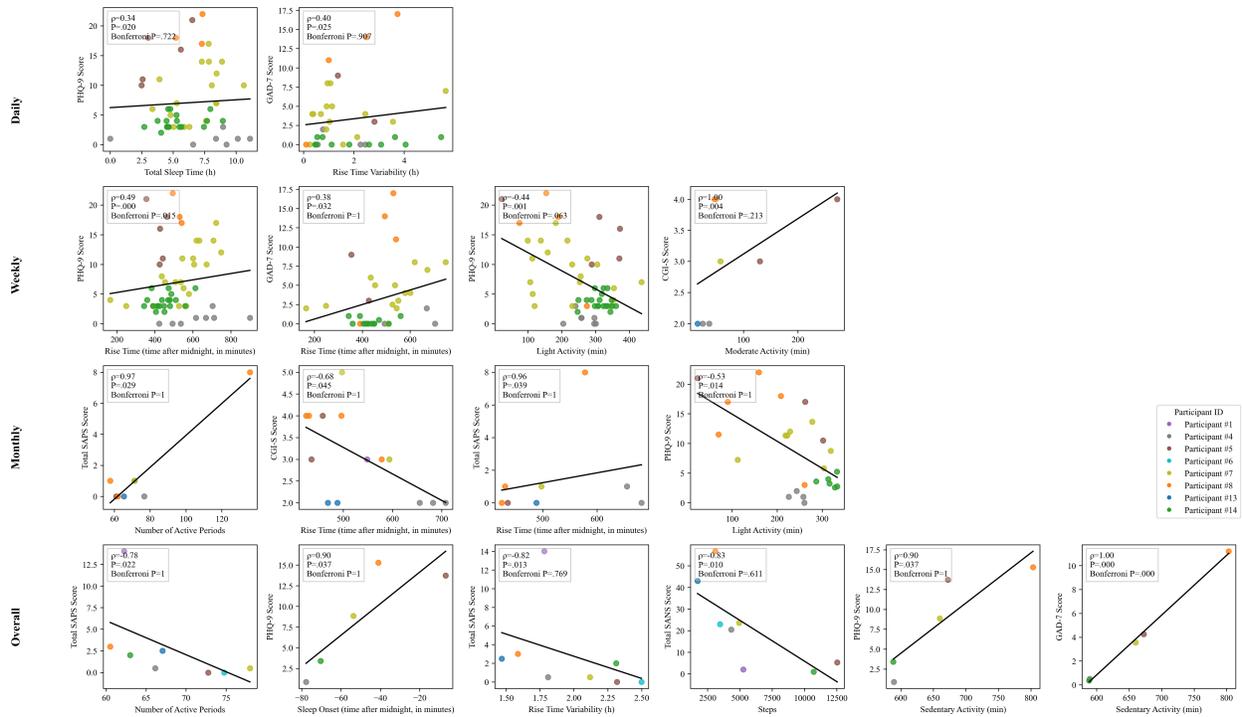

**Figure 5.** Complete Set of Significant Inter-Individual Associations Between Actigraphy Features and Questionnaire Scores Across Time Scales. Spearman correlations were used for overall averages across the full study duration. * indicates fewer than 10 data points; Spearman correlation may be less reliable.

**Table 3.** $R^2$ values of actigraphy and questionnaire scores over time for each participant.

| Variable | Participant #1 | Participant #4 | Participant #5 | Participant #6 | Participant #7 | Participant #8 | Participant #13 | Participant #14 |
|---|---|---|---|---|---|---|---|---|
| Total Sleep Time $R^2$ | 0.02 | 0.05 | 0.01 | 0.02 | 0.05 | 0.05 | 0.00 | 0.12 |
| Sleep Efficiency $R^2$ | 0.06 | 0.02 | 0.00 | 0.01 | 0.06 | 0.00 | 0.01 | 0.06 |
| Number of Active Periods $R^2$ | 0.03 | 0.01 | 0.01 | 0.00 | 0.00 | 0.01 | 0.01 | 0.03 |
| Median Activity Duration $R^2$ | 0.02 | 0.00 | 0.00 | 0.00 | 0.05 | 0.00 | 0.01 | 0.04 |
| Sleep Onset $R^2$ | 0.11 | 0.02 | 0.02 | 0.05 | 0.03 | 0.00 | 0.02 | 0.05 |
| Rise Time $R^2$ | 0.08 | 0.12 | 0.03 | 0.02 | 0.01 | 0.09 | 0.00 | 0.09 |
| Sleep Onset Variability $R^2$ | 0.00 | 0.02 | 0.00 | 0.02 | 0.00 | 0.03 | 0.06 | 0.00 |
| Rise Time Variability $R^2$ | 0.05 | 0.00 | 0.06 | 0.06 | 0.04 | 0.00 | 0.02 | 0.03 |
| Steps $R^2$ | 0.01 | 0.16 | 0.02 | 0.02 | 0.02 | 0.00 | 0.00 | 0.01 |
| Sleep Duration $R^2$ | 0.04 | 0.00 | 0.03 | 0.02 | 0.00 | 0.14 | 0.07 | 0.13 |
| Sedentary Time $R^2$ | 0.18 | 0.02 | 0.32 | 0.01 | 0.11 | 0.03 | 0.00 | 0.11 |
| Light Activity $R^2$ | 0.00 | 0.02 | 0.57 | 0.08 | 0.05 | 0.32 | 0.14 | 0.01 |
| Moderate Activity $R^2$ | 0.09 | 0.17 | 0.30 | 0.00 | 0.01 | 0.05 | 0.00 | 0.02 |
| PHQ-9 $R^2$ | — | 0.37 | 0.81 | — | 0.25 | 0.63 | — | 0.18 |
| GAD-7 $R^2$ | — | 0.62 | 0.76 | — | 0.66 | 0.25 | 0.05 | 0.01 |
| CGI-S $R^2$ | 0.00 | 0.00 | 0.59 | 0.00 | 0.71 | 0.00 | 0.05 | — |
| SAPS Total $R^2$ | 0.00 | 0.48 | 0.00 | 0.00 | 0.02 | 0.69 | 0.05 | 0.00 |
| SANS Total $R^2$ | 0.00 | 0.48 | 0.05 | 0.09 | 0.02 | 0.69 | 0.05 | 0.00 |

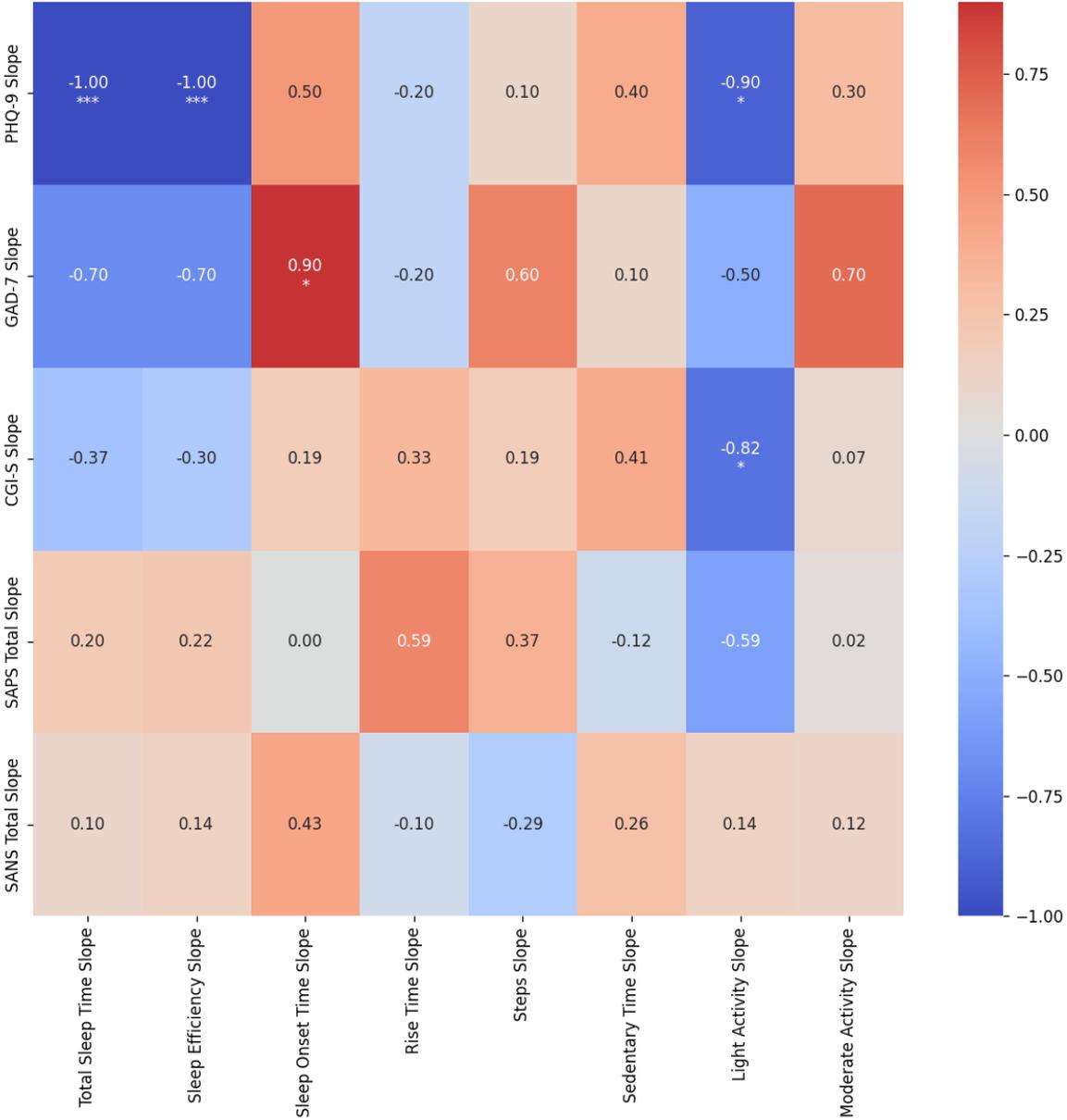

**Figure 6.** Heatmap of Spearman correlations between slopes of standardized actigraphy features and standardized clinical questionnaire scores across participants. * indicates P<.05. None of these correlations stayed significant after Bonferroni correction.